\documentclass[aps,preprint,floatfix,showpacs]{revtex4-1}
\usepackage[pdftex]{color}
\usepackage{latexsym,amssymb,amsmath,gensymb}
\usepackage{graphicx,dcolumn,bm}
\usepackage{slashed,kec}
\usepackage{mathtools}
\begin{document}

\title{Theories with Finite Green's Functions}

\author{Kevin Cahill}
\email{cahill@unm.edu}
\affiliation{Department of Physics \& Astronomy,
University of New Mexico, Albuquerque, NM 87131, USA}
\affiliation{Physics Department, Fudan University,
Shanghai 200433, China}

\date{\today}

\begin{abstract}
The addition of certain nonrenormalizable terms to the usual action density of a free scalar field leads to nonrenormalizable theories whose exact euclidian and minkowskian Green's functions are less singular than those of the free theory.
In some cases, they are finite.   One may use lattice methods to extract physical information from these less-singular,
nonrenormalizable theories. 
\end{abstract}

\maketitle

\section{Infinities\label{Infinities}}

Infinite terms have been an awkward aspect of
quantum field theory for over 80 years.
This paper will show that by adding
certain nonrenormalizable terms 
to the usual action density of a free
scalar field, one can construct nonrenormalizable theories 
whose exact euclidian and minkowskian Green's
functions are less singular than those
of the free theory.  In some cases, they are finite.
One may use lattice methods to extract physical information from these less-singular,
nonrenormalizable theories. 
The perturbative expansions of these
nonrenormalizable theories are, of course, singular.
\par
The history of attempts
to cope with the infinities of
quantum field theory is too vast
to review here, but
it may be useful for me to say
what this paper is \tit{not} about.
Most of the early work on infinities described
ways to cancel the infinities
of the perturbative expansion
of a theory against other infinite terms
present in the original lagrangian
of the same theory.  
Dimensional regularization~\cite{Hooft1972}
was a highpoint of this work.
This paper is not about 
renormalization~\cite{WeinbergI}\@.
Somewhat more-recent work 
used space-time 
lattices~\cite{PhysRevD.10.2445, *PhysRevLett.43.553}
or strong-coupling 
expansions~\cite{Parisi1978, *PhysRevD.19.1865, *PhysRevD.23.2976}.
This paper has
nothing to do with these techniques, but
one may use them to extract physical
information from the nonrenormalizable theories
to which this paper points.
Over the past three decades,
string theorists have constructed
theories that are intrinsically finite
because their basic objects are extended
in at least one 
dimension~\cite{Becker2007}\@.
This paper is much more modest.
Its main point 
is that some nonrenormalizable theories
are less singular than theories that are free or
renormalizable.  Its only antecedent, as far as 
I know, is a very interesting
paper by Boettcher and Bender~\cite{Bender1990}\@.
\par
I will discuss Green's functions first
in euclidian space and then in Minkowski space. 
 
\section{Euclidian Green's Functions
\label{Euclidian Green's Functions}}
\par
The mean value
in the ground state 
of a euclidian-time-ordered product
of fields is a ratio of path integrals~\cite{CahillXVI}
\begin{equation}
G_e(x_1,\dots,x_n) \equiv 
\langle 0 | \mathcal{T} 
\left[\phi_e(x_1) \dots \phi_e(x_n)\right] | 0 \rangle
= \frac{\displaystyle{\int}
\phi(x_1) \dots \phi(x_n)
\, \exp \left[ - \int \!
L_e(\phi) \, d^4x \right] D\phi }
{\displaystyle{\int} 
\exp \left[ - \int \!
L_e(\phi) \, d^4x \right] D\phi}
\label {euclidian time-ordered product}
\end{equation}
in which \( L_e \) is the
euclidian action density
and the time dependence
of the field is
\( \phi_e(t,\vec x) = e^{t H} \phi_e(t,\vec x)
e^{- t H} \) where \( H \) is the
hamiltonian.
If the action density is 
quadratic 
\begin{equation}
L_e = \thalf (\p_\mu \phi)^2 
+ \thalf m^2 \phi^2
= \thalf \left(\dot \phi\right)^2
+ \thalf \left(\nabla \phi\right)^2 + \thalf m^2 \phi^2 ,
\label {free action density}
\end{equation}
we can compute 
the Green's functions 
by doing gaussian integrals.
The 2-point function is
\begin{equation}
   \begin{split}
G_e(x_1,x_2) = {}& \langle 0 | \mathcal{T} 
\left[\phi_e(x_1) \phi_e(x_2)\right] | 0 \rangle 
=  \frac{\displaystyle{\int}
\phi(x_1) \phi(x_2)
\, \exp \left[ - \int \!
L_e(\phi) \, d^4x \right] D\phi }
{\displaystyle{\int} 
\exp \left[ - \int \!
L_e(\phi) \, d^4x \right] D\phi} \\
= {}& \Delta_e(x_1-x_2) =
\int \frac{e^{i p(x_1-x_2)}}{p^2+m^2} 
\frac{d^4p}{(2\pi)^4}  .
\label {2-point function}
   \end{split}
\end{equation} 
It diverges quadratically 
as \( \ep \equiv |x_1-x_2| \to 0 \)
\begin{equation}
\lim_{x_2 \to x_1}
\langle 0 | \mathcal{T} 
\left[\phi_e(x_1) \phi_e(x_2)\right] | 0 \rangle =
\langle 0 | \phi_e^2(x_1) | 0 \rangle \propto \frac{1}{\ep^2}.
\label {quadratic divergence of 2-point function}
\end{equation}
In what follows, I will show that
the addition of certain nonrenormalizable terms
to the action density (\ref{free action density})
sufficiently damps the field fluctuations 
of the resulting nonrenormalizable theory
as to make its Green's functions
less singular or even finite.

\section{Toy Theories in Euclidian Space
\label{Toy Theories in Euclidian Space}}

Toy theories without derivatives
are easy to analyze because 
their functional integrals are
infinite products of ordinary integrals.
In the toy theory
with \( L_e = m^2 \phi^2 \)
and no derivative terms,
the 2-point function
\( G_e(x,x) \) is a ratio
of products of integrals all but
one of which cancel
\begin{equation}
  \begin{split}
\langle 0 | \phi^2(x) | 0 \rangle
= {} & \frac{\displaystyle{\int}
\phi^2(x)
\, \exp \left\{ - \int \! m^2\phi^2(x') 
\, d^4x' \right\} D\phi }
{\displaystyle{\int} 
\exp \left\{ - \int \! m^2\phi^2(x') 
\, d^4x' \right\} D\phi} \\
= {} &  \frac{\displaystyle{\int}
\phi^2(x)
\, \exp \left\{ - m^2 \phi^2(x) 
\, d^4x \right\} d\phi(x) }
{\displaystyle{\int} 
\exp \left\{ - m^2 \phi^2(x) 
\, d^4x \right\} d\phi(x)}.
\label {2-point function of toy theory}
   \end{split}
\end{equation}
Setting \( d^4x = \ep^4 \) 
and \( y = m \phi(x) \ep^2 \),
we find
\begin{equation}
   \begin{split}
     \langle 0 | \phi^2(x) | 0 \rangle  = {}& 
     \lim_{\ep \to 0} \frac{1}{m^2 \ep^4}
     \frac{\displaystyle{\int}
y^2
\, e^{-y^2} \, dy }
{\displaystyle{\int} 
e^{ - y^2} \, dy} = \frac{1}{2 m^2} \, \lim_{\ep \to 0} \frac{1}{\ep^4} .
\label {E of toy theory simplified}
   \end{split}
\end{equation}
The 2-point function 
of this toy theory
without derivatives
diverges quartically.
This makes perfect sense
because if we remove
the \( p^2 \) from the
denominator of the
2-point function 
(\ref{2-point function}),
then it too diverges quartically.
The derivatives of the soluble
theory (\ref{free action density})
tether the field \( \phi(x) \)
to its values at 
neighboring points and
so reduce the divergence of 
the mean value of its square
\( \langle 0 | \phi^2(x) | 0 \rangle \) 
from quartic to quadratic.
\par
We now add a quartic 
interaction and 
consider the toy theory
with action density
\( L_e = m^2 \phi^2(x) + \lambda \phi^4(x) \)\@.
The 2-point function \( G_e(x,x) \)
is again a ratio
of products of integrals all but
one of which cancel
\begin{equation}
  \begin{split}
\langle 0 | \phi^2(x) | 0 \rangle
= {} & \frac{\displaystyle{\int}
\phi^2(x)
\, \exp \left\{ - \int \! m^2\phi^2(x')+ \lambda \phi^4(x') 
\, d^4x' \right\} D\phi }
{\displaystyle{\int} 
\exp \left\{ - \int \! m^2\phi^2(x') + \lambda \phi^4(x')
\, d^4x' \right\} D\phi} \\
= {} & \frac{\displaystyle{\int}
\phi^2(x)
\, \exp \left\{ - \left[ m^2\phi^2(x)+ \lambda \phi^4(x) \right]
\, d^4x \right\} d\phi(x) }
{\displaystyle{\int} 
\exp \left\{ - \left[ m^2\phi^2(x) + \lambda \phi^4(x) \right]
\, d^4x \right\} d\phi(x)}.
\label {E of quartic toy theory}
   \end{split}
\end{equation}
Setting \( d^4x = \ep^4 \)
and \( y = \lambda^{1/4} \ep \phi(x) \),
we have
\begin{equation}
\langle 0 | \phi^2(x) | 0 \rangle = \lim_{\ep \to 0}
\frac{1}{\sqrt{\lambda} \, \ep^2}
\frac{ \displaystyle{\int} y^2 \exp\left[ -\left( 
\ep^2 m^2 y^2/\sqrt{\lambda} + y^4
\right) \right] dy}
{\displaystyle{\int}  \exp\left[ -\left( 
\ep^2 m^2 y^2/\sqrt{\lambda} + y^4
\right)\right] dy}
= \frac{\Gamma(3/4)}{\Gamma(1/4) \sqrt{\lambda}}  \, \lim_{\ep \to 0} \frac{1}{\ep^2} 
\label {toy quartic model diverges quadratically}
\end{equation}
apart from finite terms.
The quartic term \( \lambda \phi^4(x) \) 
in the action density has provided enough damping
to reduce the divergence
of \( G_e(x,x) \) from quartic to quadratic.
\par
The action density of our third toy model is
\( L_e = m^2 \phi^2(x) 
+ \lambda \mu^{4-2n} \phi^{2n}(x) \)
in which \( n \ge 2 \), and \( \mu \) is a mass parameter.
Boettcher and Bender have studied the
\( n \to \infty \) limit of this model~\cite{Bender1990}\@.
Now after canceling identical integrals
in the ratio of path integrals and
setting \( d^4x = \ep^4 \) and
\( y = \lambda^{1/2n} \mu^{2/n-1} 
\ep^{2/n} \phi(x) \), we get
\begin{equation}
   \begin{split}
   \langle 0 | \phi^2(x) | 0 \rangle = {}& 
   \lim_{\ep \to 0}
   \frac{ \displaystyle{\int} \phi^2 (x) 
   \exp\left\{-\left[ m^2 \phi^2(x) 
   +  \lambda \mu^{4-2n} \phi^{2n}(x) 
   \right] d^4x \right\} d\phi(x) }
{\displaystyle{\int}  \exp\left\{-\left[ m^2 \phi^2(x) 
   +  \lambda \mu^{4-2n} \phi^{2n}(x) 
\right] d^4x \right\} d\phi(x) } \\
= {}&    \lim_{\ep \to 0}
\frac{\mu^{2-4/n}}{\lambda^{1/n} \, \ep^{4/n}}
\frac{ \displaystyle{\int} y^2 \exp\left\{-\left[ 
\ep^{4-4/n} m^2 \mu^{2-4/n} \lambda^{-1/n} y^2 + y^{2n}
\right]\right\} dy}
{\displaystyle{\int}  \exp\left\{-\left[ 
\ep^{4-4/n} m^2 \mu^{2-4/n} \lambda^{-1/n} y^2 + y^{2n}
\right]\right\} dy} \\
= {}& \frac{\mu^{2-4/n}}{\lambda^{1/n}} 
\frac{\Gamma(3/2n)}{\Gamma(1/2n)} \,\,
\lim_{\ep \to 0} \frac{1}{\ep^{4/n}}
\label {phi^2n model diverges softly}
   \end{split}
\end{equation}
apart from terms that are finite.
As \( n \) rises, the singularity
in the Green's function \( G_e(x,x) \)
softens.  For \( n = 4 \), the divergence
is linear; for \( n = 8 \), it is a square-root.
\par
The action density of our
fourth and final nonderivative toy theory is
\begin{equation}
L_e = 
m^2 M^2 \left( \frac{1}{1- \phi^2/M^2} -1 \right)
\equiv 
m^2 M^2 \sum_{\ell = 1}^\infty 
\frac{\phi^{2\ell}}{M^{2\ell}}  .
\label {phi inf toy theory}
\end{equation}
It is infinite for \( \phi^2 \ge M^2 \)\@.
This singularity effectively limits the path integral
to fields in the range \( - M < \phi(x) < M \)
for all space-time points \( x \)\@. 
Setting \( d^4x = \ep^4 \)
and \( y = \phi(x)/M \), we find,
after a cancellation in which only the
integration over \( \phi(x) \) 
survives, that even the \(2n\)-point function
\begin{equation}
  \begin{split}
\langle 0 | \phi^{2n}(x) | 0 \rangle
= {} & \frac{\displaystyle{\int}
\phi^{2n}(x)
\, \exp \left\{ - \left[ 
 m^2 M^2 \left( \frac{1}{1 - \phi^2(x)/M^2} -1 \right) \right]
d^4x\right\} d\phi(x)}
{\displaystyle{\int} 
\exp \left\{ - \left[ 
 m^2 M^2 \left( \frac{1}{1 - \phi^2(x)/M^2} -1 \right) \right]
d^4x \right\} d\phi(x)} \\
= {} & \lim_{\ep \to 0} \,
\frac{\displaystyle{\int_{-M}^M}
\phi^{2n}
\, \exp \left[ - \ep^4
\left( 
\frac{ m^2 M^2}{1 - \phi^2/M^2} \right)
\right] d\phi}
{\displaystyle{\int_{-M}^M} 
\exp \left[ - \ep^4
\left( 
 \frac{ m^2 M^2}{1 - \phi^2/M^2} \right)
\right] d\phi} \\
= {} & M^{2n} \, \lim_{\ep \to 0} \,
\frac{\displaystyle{\int_{-1}^1}
y^{2n} \exp\left[ - \ep^4 \left( 
\frac{ m^2 M^2}{1 - y^2} \right) \right]
\,  dy}
{\displaystyle{\int_{-1}^1} 
\exp\left[ - \ep^4 \left( 
\frac{ m^2 M^2}{1 - y^2} \right) \right] dy} \\
= {} & M^{2n} \,
\frac{\displaystyle{\int_{-1}^1}
y^{2n}
\,  dy}
{\displaystyle{\int_{-1}^1} 
 dy} = \frac{M^{2n}}{2n+1} 
\label {E of exp toy theory}
   \end{split}
\end{equation}
is finite.

\section{Lattice Models of Theories with Derivatives
in Euclidian Space
\label{Lattice Models of Theories with Derivatives in Euclidian Space}}

We now add derivative terms 
to our toy models.
The first toy model 
becomes the soluble theory with
2-point function 
(\ref{2-point function})\@.
The action density of the
second toy model with derivatives is 
\begin{equation}
L_{4} = \half \left[ \dot \phi^2 + (\nabla \phi)^2 
+ m^2 \phi^2 \right] + \lambda \phi^4  .
\label {L4e}
\end{equation}
We can put it
on a lattice of spacing \( a \)
if we take the action \( S \) to be
a sum over all vertices 
\( v \) of the vertex action
\begin{equation}
   \begin{split}
S_{4,v} = {}& \frac{a^4}{4}   
\sum_{j=1}^4 
\left(\frac{\phi(v) - \phi(v + \hat j)}{a} \right)^2
+ \frac{ a^4 m^2}{2}  \phi^2(v) 
+ a^4 \lambda \phi^4(v) \\
= {}& \fourth
\sum_{j=1}^4 
\left(\varphi(v) - \varphi(v+\hat j)\right)^2
+ \half a^2 m^2  \varphi^2(v) 
+ \lambda \varphi^4(v) 
  \label {S4n without a}
  \end{split}
\end{equation}
in which each vertex is labelled 
by four integers \( v = (n_1, n_2, n_3, n_4) \),
the field \( \varphi = a \phi \)
is dimensionless, and
\( \hat j_k = \delta_{j, k} \)\@.
Apart from the mass term,
the lattice spacing \( a \) has disappeared
from the action, but it reappears in 
the Green's functions
\begin{equation}
	 \begin{split}
\langle 0 | \mathcal{T} 
\left[\phi_e(v_1) \dots \phi_e(v_n)\right] | 0 \rangle
= {}& \frac{\displaystyle{\int}
\phi(v_1) \dots \phi(v_n)
\, \exp \left[ - \int \!
L_{4}(\phi) \, d^4x \right] D\phi }
{\displaystyle{\int} 
\exp \left[ - \int \!
L_{4}(\phi) \, d^4x \right] D\phi} \\
= {}&  \lim_{a \to 0} \,
\frac{1}{a^n} \,
\frac{\displaystyle{\int}
\varphi(v_1) \dots \varphi(v_n)
\, \exp \left( - \sum_v S_{4,v} \right) 
\prod_v d\varphi(v) }
{\displaystyle{\int} 
\exp \left( - \sum_v S_{4,v} \right) 
\prod_v d\varphi(v)} .
\label {euclidian n-point function}
   \end{split}
\end{equation}
For instance, the 2-point function is
\begin{equation}
	 \begin{split}
\langle 0 | \mathcal{T} 
\left[\phi_e(v_1) \phi_e(v_2)\right] | 0 \rangle
= {}&  \lim_{a \to 0} 
\frac{1}{a^2}
\frac{\displaystyle{\int}
\varphi(v_1) \varphi(v_2)
\, \exp \left( - \sum_v S_{4,v} \right) 
\prod_v d\varphi(v) }
{\displaystyle{\int} 
\exp \left( - \sum_v S_{4,v} \right) 
\prod_v d\varphi(v) } .
\label {euclidian 2-point function}
   \end{split}
\end{equation}
As \( v_2 \to v_1 \), 
this ratio still diverges quadratically,
like its toy twin
(\ref{toy quartic model diverges quadratically}),
so the
quartic and derivative terms don't
conspire to further
reduce this divergence in
\( G_e(v_1,v_1) \)\@.
\par
The third toy model with
derivatives has action density
\begin{equation}
L_{2n} = \thalf \left[ \dot \phi^2 + (\nabla \phi)^2 
+ m^2 \phi^2 \right] + \lambda \mu^{4-2n} \phi^{2n}  .
\label {L2nder}
\end{equation}
Boettcher and Bender have studied the
\( n \to \infty \) limit of this model~\cite{Bender1990}\@.
Its lattice action \( S \) is
a sum over all vertices \( v \) of 
\begin{equation}
   \begin{split}
S_{2n,v} = {}& \frac{a^4}{4}   
\sum_{j=1}^4 
\left(\frac{\phi(v) - \phi(v+\hat j)}{a} \right)^2
+ \frac{ a^4 m^2}{2}  \phi^2(v) 
+ a^4 \lambda \mu^{4-2n} \phi^{2n}(v) \\
= {}& 
\frac{a^{2-4/n} \lambda^{-1/n} \mu^{2-4/n}}{4}
\sum_{j=1}^4 
\left(\varphi(v) - \varphi(v+\hat j)\right)^2 \\
{}& + \half a^{4-4/n} \lambda^{-1/n} 
m^2 \mu^{2-4/n} \varphi^2(v) 
+ \varphi^{2n}(v) 
  \label {S4ell without a}
  \end{split}
\end{equation}
in which the field \( \varphi(v) = 
\lambda^{1/2n} \mu^{2/n-1} a^{2/n} \phi(v) \)
is dimensionless.
The 2-point function 
\begin{equation}
	 \begin{split}
\langle 0 | \mathcal{T} 
\left[\phi_e(v_1) \phi_e(v_2)\right] | 0 \rangle
= {}&  \lim_{a \to 0} \, 
\frac{\mu^{2-4/n}}{\lambda^{1/n}a^{4/n}} \,
\frac{\displaystyle{\int}
\varphi(v_1) \varphi(v_2)
\, \exp \left( - \sum_v S_{2n,v} \right) 
\prod_v d\varphi(v) }
{\displaystyle{\int} 
\exp \left( - \sum_v S_{2n,v} \right) 
\prod_v d\varphi(v) } 
\label {2n euclidian 2-point function}
   \end{split}
\end{equation}
for \( n > 2 \) and \( v_1 = v_2 \)
is less singular than \( 1/a^2 \)\@.
\par
The fourth toy model with derivatives is
\begin{equation}
L_e = \thalf (\p_\mu\phi)^2 
+ \thalf  m^2 M^2 
\bigg( \frac{1}{1- \phi^2/M^2} -1 \bigg)
\equiv \thalf (\p_\mu\phi)^2 
+ \thalf  m^2 M^2 \sum_{n = 1}^\infty 
\frac{\phi^{2n}}{M^{2n}} .
\label {phi inf theory}
\end{equation}
Its lattice action is a sum over all
vertices \( v \) of the vertex action
\begin{equation}
   \begin{split}
S_{M,v} = {}& \frac{a^4}{4}   
\sum_{j=1}^4 
\left(\frac{\phi(v) - \phi(v+\hat j)}{a} \right)^2
+ \frac{a^4m^2 M^2}{2} \left(
\frac{1}{1 - \phi^2(v)/M^2} -1 \right)\\
= {}& \frac{a^2M^2}{4}   
\sum_{j=1}^4 
\left( \varphi(v) - \varphi(v+\hat j) \right)^2
+ \frac{a^4 m^2 M^2}{2} 
\left( \frac{1}{1 - \varphi^2(v)} -1 \right)
\label {SMell}
    \end{split}
\end{equation}
in which the field
\( \varphi = \phi/M \) is
dimensionless.
The essential singularity in the
functional integrals effectively 
restricts the field
\( \varphi(v) \) to the interval
\( -1 < \varphi(v) < 1 \)\@.
The \(2n\)-point function 
\begin{equation}
	 \begin{split}
\langle 0 | \mathcal{T} 
\left[\phi_e(v_1) \dots \phi_e(v_{2n})\right] | 0 \rangle
= {}&  M^{2n} \, \lim_{a \to 0} \,
\frac{\displaystyle{\int_{-1}^1}
\varphi(v_1) \dots \varphi(v_{2n})
\, \exp \left( - \sum_v S_{M,v} \right) 
\prod_v d\varphi(v) }
{\displaystyle{\int_{-1}^1} 
\exp \left( - \sum_v S_{M,v} \right) 
\prod_v d\varphi(v) } 
\label {M euclidian n-point function}
   \end{split}
\end{equation}
is finite for all  \( n \),
even when all the points coincide,
\( v_j = v_0 \),\begin{equation}
	 \begin{split}
\langle 0 | 
\phi^{2n}_e(v_0) | 0 \rangle
= {}&  M^{2n} \, \lim_{a \to 0} \,
\frac{\displaystyle{\int_{-1}^1}
\varphi^{2n}(v_0) 
\, \exp \left( - \sum_v S_{M,v} \right) 
\prod_v d\varphi(v) }
{\displaystyle{\int_{-1}^1} 
\exp \left( - \sum_v S_{M,v} \right) 
\prod_v d\varphi(v) } \\
= {}&  \frac{M^{2n}}{2n+1} . 
\label {M euclidian 2n-point function}
   \end{split}
\end{equation}

\section{Green's Functions in Minkowski Space
\label{Green's Functions in Minkowski Space}}

\par
We have seen that the addition
of certain nonrenormalizable terms to 
the euclidian action density of a theory of scalar fields
can make the euclidian Green's functions 
of the resulting nonrenormalizable theory
less singular or even finite.  
To extend these results to
Green's functions in Minkowski space
and avoid extra notation,
I will continue to focus on theories of a single scalar field.
The mean value
in the ground state 
of a time-ordered product
of fields is a ratio of path integrals~\cite{CahillXVI}\begin{equation}
G(x_1,\dots,x_n) \equiv 
\langle 0 | \mathcal{T} 
\left[\phi(x_1) \dots \phi(x_n)\right] | 0 \rangle
= \frac{\displaystyle{\int}
\phi(x_1) \dots \phi(x_n)
\, \exp \left[ i \! \int \!
L(\phi) \, d^4x \right] D\phi }
{\displaystyle{\int} 
\exp \left[ i \! \int \!
L(\phi) \, d^4x \right] D\phi}
\label {time-ordered product}
\end{equation}
in which \( L \) is the
action density
and the time dependence
of the field \( \phi(x) \) is
\( \phi(t,\vec x) = e^{i t H} \phi(t,\vec x)
e^{- i t H} \) where \( H \) is the
hamiltonian.
The symbol \( D\phi \) means that
we should integrate over all real functions
\( \phi(x) \) of space-time and also should
include in both the numerator and the
denominator the factors
\( \la 0 | \phi(\infty,\vec x) \ra \) 
and \( \la \phi(- \infty,\vec x) | 0 \ra \),
which lead to the \( i \epsilon \)
terms in propagators.
Green's functions play a central role
in quantum field theory; they occur, for example,
in the LSZ reduction formula~\cite{SrednickiV}
for the scattering of \( n \) incoming
particles of momenta \( p_1 \dots p_n \equiv \{p\} \)
into \( n' \) outgoing
particles of momenta \( p'_1 \dots p'_n \equiv \{p'\} \) 
\begin{equation}
   \begin{split}
      \la p' | p \ra = {}& \prod_{\ell=1}^n \!
      \prod_{\ell'=1}^{n'} \!
      \int \! d^4x_\ell d^4x'_{\ell'} 
      e^{ip_\ell x_\ell -ip'_{\ell'} x'_{\ell'}}
      (- \p^2_\ell + m^2)       (- \p'^2_\ell + m^2) 
       \langle 0 | \mathcal{T} 
       \left[\phi(x_1) \dots \phi(x_{n+n'})\right] | 0 \rangle .
         \label {LSZ}
   \end{split}
\end{equation}
\par
If the action density is the
quadratic form
\begin{equation}
L = - \thalf \p_\mu \phi \p^\mu \phi
- \thalf m^2 \phi^2
= \thalf \left(\dot \phi\right)^2
- \thalf \left(\nabla \phi\right)^2 - \thalf m^2 \phi^2 ,
\label {free action density}
\end{equation}
then we can compute all the 
Green's functions.
The 2-point function, for instance, is
\begin{equation}
   \begin{split}
G(x_1,x_2) = {}& 
\langle 0 | \mathcal{T} 
\left[\phi(x_1) \phi(x_2)\right] | 0 \rangle \\
= {}& \frac{\displaystyle{\int}
\phi(x_1) \phi(x_2)
\, \exp \left[ i \! \int \!
L(\phi) \, d^4x \right] D\phi }
{\displaystyle{\int} 
\exp \left[ i \! \int \!
L(\phi) \, d^4x \right] D\phi} \\
= {}& \Delta(x_1-x_2) =
\int \frac{e^{i p(x_1-x_2)}}{p^2+m^2 - i \epsilon} 
\frac{d^4p}{(2\pi)^4} .
\label {2-point function}
   \end{split}
\end{equation}
It diverges quadratically 
as \( \ep \equiv |x_1-x_2| \to 0 \) 
\begin{equation}
\lim_{x_2 \to x_1}
\langle 0 | \mathcal{T} 
\left[\phi(x_1) \phi(x_2)\right] | 0 \rangle =
\langle 0 | \phi^2(x_1) | 0 \rangle \propto \frac{1}{\ep^2} .
\label {quadratic divergence of 2-point function}
\end{equation}
In what follows, I will show that
the addition of certain nonrenormalizable terms
to the action density (\ref{free action density})
makes its Green's functions (\ref{time-ordered product})
less singular or even finite.

\section{Toy Theories in Minkowski Space
\label{Toy Theories in Minkowski Space}}

In the toy theory
with \( L = - m^2 \phi^2 \)
and no derivative terms,
the 2-point function
\( G(x,x) \) is a ratio
of products of integrals all but
one of which cancel
\begin{equation}
  \begin{split}
\langle 0 | \phi^2(x) | 0 \rangle
= {} & \frac{\displaystyle{\int}
\phi^2(x)
\, \exp \left\{ - i \! \int \! m^2\phi^2(x') 
\, d^4x' \right\} D\phi }
{\displaystyle{\int} 
\exp \left\{ - i \! \int \! m^2\phi^2(x') 
\, d^4x' \right\} D\phi} \\
= {} &  \frac{\displaystyle{\int}
\phi^2(x)
\, \exp \left\{ - i m^2 \phi^2(x) 
\, d^4x \right\} d\phi(x) }
{\displaystyle{\int} 
\exp \left\{ - i m^2 \phi^2(x) 
\, d^4x \right\} d\phi(x)}.
\label {2-point function of toy theory}
   \end{split}
\end{equation}
Setting \( d^4x = \ep^4 \) 
and \( y = m \phi(x) \ep^{2} \),
we find
\begin{equation}
   \begin{split}
     \langle 0 | \phi^2(x) | 0 \rangle  = {}& 
     \lim_{\ep \to 0} \, \frac{1}{m^2 \ep^4} \,
     \frac{\displaystyle{\int_0^{\infty}}
y^2
\, e^{- i y^2} \, dy }
{\displaystyle{\int_0^{\infty}} 
e^{ - i y^2} \, dy} = 
\frac{2\sqrt{2}}{(1-i)\sqrt{\pi} \, m^2} \, 
\lim_{\ep \to 0} \, \frac{1}{\ep^4} \,
\displaystyle{\int_0^{\infty}}
y^2
\, e^{- i y^2} \, dy 
\label {E of toy theory simplified}
   \end{split}
\end{equation}
in which the final integral does not
converge.
The 2-point function 
of this toy theory
without derivatives
diverges a bit worse than quartically.
\par
We now add a quartic 
interaction and 
consider the toy theory
with action density
\( L = - m^2 \phi^2(x) - \lambda \phi^4(x) \)\@.
The 2-point function \( G(x,x) \)
is again a ratio
of products of integrals all but
one of which cancel
\begin{equation}
  \begin{split}
\langle 0 | \phi^2(x) | 0 \rangle
= {} & \frac{\displaystyle{\int}
\phi^2(x)
\, \exp \left\{ - i \int \! m^2\phi^2(x')+ \lambda \phi^4(x') 
\, d^4x' \right\} D\phi }
{\displaystyle{\int} 
\exp \left\{ - i \int \! m^2\phi^2(x') + \lambda \phi^4(x')
\, d^4x' \right\} D\phi} \\
= {} & \frac{\displaystyle{\int}
\phi^2(x)
\, \exp \left\{ - i \left[ m^2\phi^2(x)+ \lambda \phi^4(x) \right]
\, d^4x \right\} d\phi(x) }
{\displaystyle{\int} 
\exp \left\{ - i \left[ m^2\phi^2(x) + \lambda \phi^4(x) \right]
\, d^4x \right\} d\phi(x)}.
\label {E of quartic toy theory}
   \end{split}
\end{equation}
Setting \( d^4x = \ep^4 \)
and \( y = \lambda^{1/4} \ep \phi(x) \),
we have
\begin{equation}
\langle 0 | \phi^2(x) | 0 \rangle = \lim_{\ep \to 0} \,
\frac{1}{\sqrt{\lambda} \, \ep^2} \,
\frac{ \displaystyle{\int} y^2 \exp\left[ - i \left( 
\ep^2 m^2 y^2/\sqrt{\lambda} + y^4
\right) \right] dy}
{\displaystyle{\int}  \exp\left[ - i \left( 
\ep^2 m^2 y^2/\sqrt{\lambda} + y^4
\right)\right] dy}
= \frac{\Gamma(3/4)}
{(-1)^{1/4} \Gamma(1/4) \sqrt{\lambda}}  \, 
\lim_{\ep \to 0} \, \frac{1}{\ep^2} 
\label {toy quartic model diverges quadratically}
\end{equation}
or 
\( 0.238994 (1,-i) /\sqrt{\lambda} \epsilon^2 \)
apart from finite terms.
The quartic term \( - \lambda \phi^4(x) \) 
in the action density has reduced the divergence
of \( G(x,x) \) from a little more than quartic
to quadratic.
\par
The action density of our third toy model is
\( L = - m^2 \phi^2(x) 
- \lambda \mu^{4-2n} \phi^{2n}(x) \)
in which \( n \ge 2 \) and \( \mu \) is a mass parameter.
Boettcher and Bender have studied the
\( n \to \infty \) limit of this model~\cite{Bender1990}\@.
Now after canceling identical integrals
in the ratio of path integrals and
setting \( d^4x = \ep^4 \) and
\( y = \lambda^{1/2n} \mu^{2/n-1} 
\ep^{2/n} \phi(x) \), we get
\begin{equation}
   \begin{split}
   \langle 0 | \phi^2(x) | 0 \rangle = {}& 
   \lim_{\ep \to 0} \,
   \frac{ \displaystyle{\int} \phi^2 (x) 
   \exp\left\{- i \left[ m^2 \phi^2(x) 
   +  \lambda \mu^{4-2n} \phi^{2n}(x) 
   \right] d^4x \right\} d\phi(x) }
{\displaystyle{\int}  \exp\left\{- i \left[ m^2 \phi^2(x) 
   +  \lambda \mu^{4-2n} \phi^{2n}(x) 
\right] d^4x \right\} d\phi(x) } \\
= {}&    \lim_{\ep \to 0} \,
\frac{\mu^{2-4/n}}{\lambda^{1/n} \, \ep^{4/n}} \,
\frac{ \displaystyle{\int} y^2 \exp\left\{- i \left[ 
\ep^{4-4/n} m^2 \mu^{2-4/n} \lambda^{-1/n} y^2 + y^{2n}
\right]\right\} dy}
{\displaystyle{\int}  \exp\left\{- i \left[ 
\ep^{4-4/n} m^2 \mu^{2-4/n} \lambda^{-1/n} y^2 + y^{2n}
\right]\right\} dy} \\
= {}& e^{-i\pi/2n} \, \frac{\mu^{2-4/n}}{3\lambda^{1/n}} \,
\frac{\Gamma(1+ 3/2n)}{\Gamma(1+ 1/2n)} \,\,
\lim_{\ep \to 0} \, \frac{1}{\ep^{4/n}}
\label {phi^2n model diverges softly}
   \end{split}
\end{equation}
apart from terms that are finite.
As \( n \) rises, the singularity
in the Green's function \( G(x,x) \)
softens.  For \( n = 4 \), the divergence
is linear; for \( n = 8 \), it is a square-root.
\par
The action density of our 
fourth and final nonderivative toy theory is
\begin{equation}
L = -
m^2 M^2 \left( \frac{1}{1- \phi^2/M^2} -1 \right)
\equiv -
m^2 M^2 \sum_{\ell = 1}^\infty 
\frac{\phi^{2\ell}}{M^{2\ell}}  .
\label {phi inf toy theory}
\end{equation}
It is infinite for \( \phi^2 \ge M^2 \)\@.
This singularity effectively limits
the path integral to fields in the range
\( -M < \phi(x) < M \) for all space-time points \( x \)\@.
Setting \( d^4x = \ep^4 \)
and \( y = \phi(x)/M \), we find,
after a cancellation in which only the
integration over \( \phi(x) \) 
survives, that even the \( 2n \)-point function
\begin{equation}
  \begin{split}
\langle 0 | \phi^{2n}(x) | 0 \rangle
= {} & \frac{\displaystyle{\int}
\phi^{2n}(x)
\, \exp \left\{ - i \int \! 
\left[ 
 m^2 M^2 \left( \frac{1}{1 - \phi^2(x')/M^2} -1 \right) \right] 
d^4x' \right\} D\phi }
{\displaystyle{\int} 
\exp \left\{ - i \int \! \left[
 m^2 M^2 \left( \frac{1}{1 - \phi^2(x')/M^2} -1 \right) \right]
d^4x' \right\} D\phi} \\= {} & \frac{\displaystyle{\int}
\phi^{2n}(x)
\, \exp \left\{ - i \left[ 
 m^2 M^2 \left( \frac{1}{1 - \phi^2(x)/M^2} -1 \right) \right]
d^4x\right\} d\phi(x)}
{\displaystyle{\int} 
\exp \left\{ - i \left[ 
 m^2 M^2 \left( \frac{1}{1 - \phi^2(x)/M^2} -1 \right) \right]
d^4x \right\} d\phi(x)} \\
= {} & \lim_{\ep \to 0} \,
\frac{\displaystyle{\int_{-M}^M}
\phi^{2n}
\, \exp \left[ - i \ep^4
\left( 
\frac{ m^2 M^2}{1 - \phi^2/M^2} \right)
\right] d\phi}
{\displaystyle{\int_{-M}^M} 
\exp \left[ - i \ep^4
\left( 
 \frac{ m^2 M^2}{1 - \phi^2/M^2} \right)
\right] d\phi} \\
= {} & M^{2n} \, \lim_{\ep \to 0} \,
\frac{\displaystyle{\int_{-1}^1}
y^{2n} \exp\left[ - i \ep^4 \left( 
\frac{ m^2 M^2}{1 - y^2} \right) \right]
\,  dy}
{\displaystyle{\int_{-1}^1} 
\exp\left[ - i \ep^4 \left( 
\frac{ m^2 M^2}{1 - y^2} \right) \right] dy} \\
= {} & M^{2n} \,
\frac{\displaystyle{\int_{-1}^1}
y^{2n}
\,  dy}
{\displaystyle{\int_{-1}^1} 
 dy} = \frac{M^{2n}}{2n+1} 
\label {E of exp toy theory}
   \end{split}
\end{equation}
is finite.
By symmetry, one has 
\( \la 0 | \phi^{2n+1}(x) | 0 \ra = 0 \)\@.  

\section{Lattice Models of Theories with Derivatives in Minkowski Space
\label{Lattice Models of Theories with Derivatives in Minkowski Space}}

We now add derivative terms 
to our toy models.
The first toy model 
becomes the soluble theory with
2-point function 
(\ref{2-point function})\@.
The action density of the
second toy model with derivatives is 
\begin{equation}
L_{4} = \half \left[ \dot \phi^2 - (\nabla \phi)^2 
- m^2 \phi^2 \right] - \lambda \phi^4  .
\label {L4e}
\end{equation}
We can put it
on a lattice of spacing \( a \)
if we take the action \( S \) to be
a sum over all vertices 
\( v \) of the vertex action
\begin{equation}
   \begin{split}
S_{4,v} = {}& - \frac{a^4}{4}   
\sum_{j=0}^3 \eta^{jj}
\left(\frac{\phi(v) - \phi(v + \hat j)}{a} \right)^2
- \frac{ a^4 m^2}{2}  \phi^2(v) 
- a^4 \lambda \phi^4(v) \\
= {}& - \fourth
\sum_{j=0}^3 \eta^{jj}
\left(\varphi(v) - \varphi(v+\hat j)\right)^2
- \half a^2 m^2  \varphi^2(v) 
- \lambda \varphi^4(v) 
  \label {S4n without a}
  \end{split}
\end{equation}
in which each vertex is labelled 
by four integers \( v = (n_0, n_1, n_2, n_3) \),
the field 
\( \varphi = a \phi \)
is dimensionless,
\( \hat j_k = \delta_{j, k} \),
and \( \eta \) is the diagonal metric
of flat space \( \mbox{diag} ( \eta ) = (-1, 1, 1, 1) \)\@.
Apart from the mass term,
the lattice spacing \( a \) has disappeared
from the action, but it reappears in 
the \(n\)-point functions
\begin{equation}
	 \begin{split}
\langle 0 | \mathcal{T} 
\left[\phi(v_1) \dots \phi(v_n)\right] | 0 \rangle
= {}& \frac{\displaystyle{\int}
\phi(v_1) \dots \phi(v_n)
\, \exp \left[ - i \! \int \!
L_{4}(\phi) \, d^4x \right] D\phi }
{\displaystyle{\int} 
\exp \left[ - i \! \int \!
L_{4}(\phi) \, d^4x \right] D\phi} \\
= {}&  \lim_{a \to 0} \,
\frac{1}{a^n} \,
\frac{\displaystyle{\int}
\varphi(v_1) \dots \varphi(v_n)
\, \exp \left( - i \sum_v S_{4,v} \right) 
\prod_v d\varphi(v) }
{\displaystyle{\int} 
\exp \left( - i \sum_v S_{4,v} \right) 
\prod_v d\varphi(v)} .
\label { n-point function}
   \end{split}
\end{equation}
For instance, the 2-point function is
\begin{equation}
	 \begin{split}
\langle 0 | \mathcal{T} 
\left[\phi(v_1) \phi(v_2)\right] | 0 \rangle
= {}&  \lim_{a \to 0} \,
\frac{1}{a^2} \,
\frac{\displaystyle{\int}
\varphi(v_1) \varphi(v_2)
\, \exp \left( - i \sum_v S_{4,v} \right) 
\prod_v d\varphi(v) }
{\displaystyle{\int} 
\exp \left( - i \sum_v S_{4,v} \right) 
\prod_v d\varphi(v) } .
\label { 2-point function}
   \end{split}
\end{equation}
When \( v_1 = v_2 \), 
this ratio still diverges quadratically,
like its toy twin
(\ref{toy quartic model diverges quadratically}),
so the
quartic and derivative terms don't
conspire to further
reduce this divergence in
\( G(x,x) \)\@.
\par
The third toy model with
derivatives has action density
\begin{equation}
L_{2n} = \half \left[ \dot \phi^2 - (\nabla \phi)^2 
- m^2 \phi^2 \right] - \lambda \mu^{4-2n} \phi^{2n}  .
\label {L2nder}
\end{equation}
Boettcher and Bender have studied the
\( n \to \infty \) limit of this model~\cite{Bender1990}\@.
Its lattice action \( S \) is
a sum over all vertices \( v \) of 
\begin{equation}
   \begin{split}
S_{2n,v} = {}& - \frac{a^4}{4}   
\sum_{j=0}^3 \eta^{j j}
\left(\frac{\phi(v) - \phi(v+\hat j)}{a} \right)^2
- \frac{ a^4 m^2}{2}  \phi^2(v) 
- a^4 \lambda \mu^{4-2n} \phi^{2n}(v) \\
= {}& -
\frac{a^{2-4/n} \lambda^{-1/n} \mu^{2-4/n}}{4}
\sum_{j=0}^3 \eta^{j j} 
\left(\varphi(v) - \varphi(v+\hat j)\right)^2 \\
{}& - \half a^{4-4/n} \lambda^{-1/n} 
m^2 \mu^{2-4/n} \varphi^2(v) 
- \varphi^{2n}(v) 
  \label {S4ell without a}
  \end{split}
\end{equation}
in which the 
field \( \varphi(v) = 
\lambda^{1/2n} \mu^{2/n-1} a^{2/n} \phi(v) \)
is dimensionless.
The 2-point function 
\begin{equation}
	 \begin{split}
\langle 0 | \mathcal{T} 
\left[\phi(v_1) \phi(v_2)\right] | 0 \rangle
= {}&  \lim_{a \to 0} \,
\frac{\mu^{2-4/n}}{\lambda^{1/n}a^{4/n}} \,
\frac{\displaystyle{\int}
\varphi(v_1) \varphi(v_2)
\, \exp \left( - i \sum_v S_{2n,v} \right) 
\prod_v d\varphi(v) }
{\displaystyle{\int} 
\exp \left( - i \sum_v S_{2n,v} \right) 
\prod_v d\varphi(v) } 
\label {2n  2-point function}
   \end{split}
\end{equation}
for \( n > 2 \) and \( v_1 = v_2 \)
is less singular than \( 1/a^2 \)\@.
\par
The fourth toy model with derivatives is
\begin{equation}
L = - \half \p_\mu\phi \, \p^\mu \phi
- \half  m^2 M^2 
\bigg( \frac{1}{1- \phi^2/M^2} -1 \bigg)
\equiv - \half \p_\mu\phi \, \p^\mu \phi
- \half  m^2 M^2 \sum_{n = 1}^\infty 
\frac{\phi^{2n}}{M^{2n}} .
\label {phi inf theory}
\end{equation}
Its lattice action is a sum over all
vertices \( v \) of the vertex action
\begin{equation}
   \begin{split}
S_{M,v} = {}& - \frac{a^4}{4}   
\sum_{j=0}^3 \eta^{j j} 
\left(\frac{\phi(v) - \phi(v+\hat j)}{a} \right)^2
- \frac{a^4m^2 M^2}{2} \left(
\frac{1}{1 - \phi^2(v)/M^2} -1 \right)\\
= {}& - \frac{a^2M^2}{4}   
\sum_{j=0}^3 \eta^{j j}
\left( \varphi(v) - \varphi(v+\hat j) \right)^2
- \frac{a^4 m^2 M^2}{2} 
\left( \frac{1}{1 - \varphi^2(v)} -1 \right)
\label {SMell}
    \end{split}
\end{equation}
in which the field
\( \varphi = \phi/M \) is
dimensionless.
The essential singularity in the
functional integrals effectively 
restricts the field
\( \varphi(v) \) to the interval
\( -1 < \varphi(v) < 1 \)\@.
The \(2n\)-point function 
\begin{equation}
	 \begin{split}
\langle 0 | \mathcal{T} 
\left[\phi(v_1) \dots \phi(v_{2n})\right] | 0 \rangle
= {}&  M^{2n} \, \lim_{a \to 0} \,
\frac{\displaystyle{\int_{-1}^1}
\varphi(v_1) \dots \varphi(v_{2n})
\, \exp \left( - i \sum_v S_{M,v} \right) 
\prod_v d\varphi(v) }
{\displaystyle{\int_{-1}^1} 
\exp \left( - i \sum_v S_{M,v} \right) 
\prod_v d\varphi(v) } 
\label {M  n-point function}
   \end{split}
\end{equation}
is finite for all  \( n \),
even when all the points coincide,
\( v_j = v_0 \),\begin{equation}
	 \begin{split}
\langle 0 | 
\phi^{2n}(v_0) | 0 \rangle
= {}&  M^{2n} \, \lim_{a \to 0} \,
\frac{\displaystyle{\int_{-1}^1}
\varphi^{2n}(v_0) 
\, \exp \left( - i \sum_v S_{M,v} \right) 
\prod_v d\varphi(v) }
{\displaystyle{\int_{-1}^1} 
\exp \left( - i \sum_v S_{M,v} \right) 
\prod_v d\varphi(v) } \\
= {}&  \frac{M^{2n}}{2n+1} . 
\label {M  2n-point function}
   \end{split}
\end{equation}

\section{Conclusion
\label{Conclusion}}

The addition of terms like 
\( \phi^{2n} \) for \( n > 2 \) or 
\(  \lt( m^2 M^2 / 2 \rt)
\left[(1 - \phi^2/M^2)^{-1} -1 \right] \)
to the usual action density
(\ref{free action density}) of a scalar field 
leads to nonrenormalizable theories
whose exact Green's functions in 
euclidian and Minkowski space
are less singular than those of the free theory.
In some cases, they are finite.
One may use lattice methods to extract physical information from these less-singular,
nonrenormalizable theories. 
\par
If the results 
of this paper can be extended
to fields of higher spin, then 
nonrenormalizable theories
may have more to teach us,
and their lessons may be important
because of the nonrenormalizability
of general relativity
and the absence of experimental
evidence for supersymmetry~\cite{Shifman2012}\@.

\begin{acknowledgments}
I am grateful to Carl Bender, Fred Cooper, Roy Glauber,
Gary Herling, Sudhakar Prasad, Sally Seidel,
James Thomas, and especially to Franco Giuliani
and David Waxman for helpful conversations.\end{acknowledgments}
\bibliography{physics}

\begin{thebibliography}{12}%
\makeatletter
\providecommand \@ifxundefined [1]{%
 \@ifx{#1\undefined}
}%
\providecommand \@ifnum [1]{%
 \ifnum #1\expandafter \@firstoftwo
 \else \expandafter \@secondoftwo
 \fi
}%
\providecommand \@ifx [1]{%
 \ifx #1\expandafter \@firstoftwo
 \else \expandafter \@secondoftwo
 \fi
}%
\providecommand \natexlab [1]{#1}%
\providecommand \enquote  [1]{``#1''}%
\providecommand \bibnamefont  [1]{#1}%
\providecommand \bibfnamefont [1]{#1}%
\providecommand \citenamefont [1]{#1}%
\providecommand \href@noop [0]{\@secondoftwo}%
\providecommand \href [0]{\begingroup \@sanitize@url \@href}%
\providecommand \@href[1]{\@@startlink{#1}\@@href}%
\providecommand \@@href[1]{\endgroup#1\@@endlink}%
\providecommand \@sanitize@url [0]{\catcode `\\12\catcode `\$12\catcode
  `\&12\catcode `\#12\catcode `\^12\catcode `\_12\catcode `\%12\relax}%
\providecommand \@@startlink[1]{}%
\providecommand \@@endlink[0]{}%
\providecommand \url  [0]{\begingroup\@sanitize@url \@url }%
\providecommand \@url [1]{\endgroup\@href {#1}{\urlprefix }}%
\providecommand \urlprefix  [0]{URL }%
\providecommand \Eprint [0]{\href }%
\providecommand \doibase [0]{http://dx.doi.org/}%
\providecommand \selectlanguage [0]{\@gobble}%
\providecommand \bibinfo  [0]{\@secondoftwo}%
\providecommand \bibfield  [0]{\@secondoftwo}%
\providecommand \translation [1]{[#1]}%
\providecommand \BibitemOpen [0]{}%
\providecommand \bibitemStop [0]{}%
\providecommand \bibitemNoStop [0]{.\EOS\space}%
\providecommand \EOS [0]{\spacefactor3000\relax}%
\providecommand \BibitemShut  [1]{\csname bibitem#1\endcsname}%
\let\auto@bib@innerbib\@empty
\bibitem [{\citenamefont {'t~Hooft}\ and\ \citenamefont
  {Veltman}(1972)}]{Hooft1972}%
  \BibitemOpen
  \bibfield  {author} {\bibinfo {author} {\bibfnamefont {G.}~\bibnamefont
  {'t~Hooft}}\ and\ \bibinfo {author} {\bibfnamefont {M.}~\bibnamefont
  {Veltman}},\ }\href@noop {} {\bibfield  {journal} {\bibinfo  {journal}
  {Nucl.\ Phys.}\ }\textbf {\bibinfo {volume} {B44}},\ \bibinfo {pages} {189}
  (\bibinfo {year} {1972})}\BibitemShut {NoStop}%
\bibitem [{\citenamefont {Weinberg}(1995)}]{WeinbergI}%
  \BibitemOpen
  \bibfield  {author} {\bibinfo {author} {\bibfnamefont {S.}~\bibnamefont
  {Weinberg}},\ }\href@noop {} {\emph {\bibinfo {title} {The Quantum Theory of
  Fields}}},\ Vol.\ \bibinfo {volume} {I Foundations}\ (\bibinfo  {publisher}
  {Cambridge University Press},\ \bibinfo {year} {1995})\BibitemShut {NoStop}%
\bibitem [{\citenamefont {Wilson}(1974)}]{PhysRevD.10.2445}%
  \BibitemOpen
  \bibfield  {author} {\bibinfo {author} {\bibfnamefont {K.~G.}\ \bibnamefont
  {Wilson}},\ }\href {\doibase 10.1103/PhysRevD.10.2445} {\bibfield  {journal}
  {\bibinfo  {journal} {Phys. Rev. D}\ }\textbf {\bibinfo {volume} {10}},\
  \bibinfo {pages} {2445} (\bibinfo {year} {1974})}\BibitemShut {NoStop}%
\bibitem [{\citenamefont {Creutz}(1979)}]{PhysRevLett.43.553}%
  \BibitemOpen
  \bibfield  {author} {\bibinfo {author} {\bibfnamefont {M.}~\bibnamefont
  {Creutz}},\ }\href {\doibase 10.1103/PhysRevLett.43.553} {\bibfield
  {journal} {\bibinfo  {journal} {Phys. Rev. Lett.}\ }\textbf {\bibinfo
  {volume} {43}},\ \bibinfo {pages} {553} (\bibinfo {year} {1979})}\BibitemShut
  {NoStop}%
\bibitem [{\citenamefont {Benzi}\ \emph {et~al.}(1978)\citenamefont {Benzi},
  \citenamefont {Martinelli},\ and\ \citenamefont {Parisi}}]{Parisi1978}%
  \BibitemOpen
  \bibfield  {author} {\bibinfo {author} {\bibfnamefont {R.}~\bibnamefont
  {Benzi}}, \bibinfo {author} {\bibfnamefont {G.}~\bibnamefont {Martinelli}}, \
  and\ \bibinfo {author} {\bibfnamefont {G.}~\bibnamefont {Parisi}},\
  }\href@noop {} {\bibfield  {journal} {\bibinfo  {journal} {Nuclear Physics}\
  }\textbf {\bibinfo {volume} {B135(3)}},\ \bibinfo {pages} {429} (\bibinfo
  {year} {1978})}\BibitemShut {NoStop}%
\bibitem [{\citenamefont {Bender}\ \emph {et~al.}(1979)\citenamefont {Bender},
  \citenamefont {Cooper}, \citenamefont {Guralnik},\ and\ \citenamefont
  {Sharp}}]{PhysRevD.19.1865}%
  \BibitemOpen
  \bibfield  {author} {\bibinfo {author} {\bibfnamefont {C.~M.}\ \bibnamefont
  {Bender}}, \bibinfo {author} {\bibfnamefont {F.}~\bibnamefont {Cooper}},
  \bibinfo {author} {\bibfnamefont {G.~S.}\ \bibnamefont {Guralnik}}, \ and\
  \bibinfo {author} {\bibfnamefont {D.~H.}\ \bibnamefont {Sharp}},\ }\href
  {\doibase 10.1103/PhysRevD.19.1865} {\bibfield  {journal} {\bibinfo
  {journal} {Phys. Rev. D}\ }\textbf {\bibinfo {volume} {19}},\ \bibinfo
  {pages} {1865} (\bibinfo {year} {1979})}\BibitemShut {NoStop}%
\bibitem [{\citenamefont {Bender}\ \emph {et~al.}(1981)\citenamefont {Bender},
  \citenamefont {Cooper}, \citenamefont {Guralnik}, \citenamefont {Roskies},\
  and\ \citenamefont {Sharp}}]{PhysRevD.23.2976}%
  \BibitemOpen
  \bibfield  {author} {\bibinfo {author} {\bibfnamefont {C.~M.}\ \bibnamefont
  {Bender}}, \bibinfo {author} {\bibfnamefont {F.}~\bibnamefont {Cooper}},
  \bibinfo {author} {\bibfnamefont {G.~S.}\ \bibnamefont {Guralnik}}, \bibinfo
  {author} {\bibfnamefont {R.}~\bibnamefont {Roskies}}, \ and\ \bibinfo
  {author} {\bibfnamefont {D.~H.}\ \bibnamefont {Sharp}},\ }\href {\doibase
  10.1103/PhysRevD.23.2976} {\bibfield  {journal} {\bibinfo  {journal} {Phys.
  Rev. D}\ }\textbf {\bibinfo {volume} {23}},\ \bibinfo {pages} {2976}
  (\bibinfo {year} {1981})}\BibitemShut {NoStop}%
\bibitem [{\citenamefont {Becker}\ \emph {et~al.}(2007)\citenamefont {Becker},
  \citenamefont {Becker},\ and\ \citenamefont {Schwarz}}]{Becker2007}%
  \BibitemOpen
  \bibfield  {author} {\bibinfo {author} {\bibfnamefont {K.}~\bibnamefont
  {Becker}}, \bibinfo {author} {\bibfnamefont {M.}~\bibnamefont {Becker}}, \
  and\ \bibinfo {author} {\bibfnamefont {J.~H.}\ \bibnamefont {Schwarz}},\
  }\href@noop {} {\emph {\bibinfo {title} {String Theory and M-Theory}}}\
  (\bibinfo  {publisher} {Cambridge University Press},\ \bibinfo {year}
  {2007})\BibitemShut {NoStop}%
\bibitem [{\citenamefont {Boettcher}\ and\ \citenamefont
  {Bender}(1990)}]{Bender1990}%
  \BibitemOpen
  \bibfield  {author} {\bibinfo {author} {\bibfnamefont {S.}~\bibnamefont
  {Boettcher}}\ and\ \bibinfo {author} {\bibfnamefont {C.}~\bibnamefont
  {Bender}},\ }\href@noop {} {\bibfield  {journal} {\bibinfo  {journal} {J.
  Math. Phys.}\ }\textbf {\bibinfo {volume} {31(11)}},\ \bibinfo {pages} {2579}
  (\bibinfo {year} {1990})}\BibitemShut {NoStop}%
\bibitem [{\citenamefont {Cahill}(ress)}]{CahillXVI}%
  \BibitemOpen
  \bibfield  {author} {\bibinfo {author} {\bibfnamefont {K.}~\bibnamefont
  {Cahill}},\ }\href@noop {} {\emph {\bibinfo {title} {\textit{Physical
  Mathematics}}}}\ (\bibinfo  {publisher} {Cambridge University Press},\
  \bibinfo {year} {in press})\ Chap.~\bibinfo {chapter} {16}\BibitemShut
  {NoStop}%
\bibitem [{\citenamefont {Srednicki}(2007)}]{SrednickiV}%
  \BibitemOpen
  \bibfield  {author} {\bibinfo {author} {\bibfnamefont {M.}~\bibnamefont
  {Srednicki}},\ }\href@noop {} {\emph {\bibinfo {title} {Quantum Field
  Theory}}}\ (\bibinfo  {publisher} {Cambridge University Press},\ \bibinfo
  {address} {Cambridge},\ \bibinfo {year} {2007})\ Chap.~\bibinfo {chapter}
  {5}\BibitemShut {NoStop}%
\bibitem [{\citenamefont {Shifman}(2012)}]{Shifman2012}%
  \BibitemOpen
  \bibfield  {author} {\bibinfo {author} {\bibfnamefont {M.}~\bibnamefont
  {Shifman}},\ }\href@noop {} {\  (\bibinfo {year} {2012})},\ \Eprint
  {http://arxiv.org/abs/1211.0004} {arXiv:1211.0004 [physics.pop-ph]}
  \BibitemShut {NoStop}%
\end{thebibliography}%
\end{document}